
\magnification=\magstep1
\vsize 23 true cm
\hsize 16 true cm
\def\\{\hfil\break}
\parindent 20pt
\parskip 8pt
\baselineskip 20pt
\centerline{{\bf BLACK HOLES CANNOT SUPPORT CONFORMAL SCALAR HAIR}}
\vskip 3 true cm
\centerline{T. Zannias\footnote{*}{PACS 04.20.Ex,97.60Sm}}
\centerline{Department of Physics}
\centerline{Queen's University}
\centerline{Kingston, Ontario}
\centerline{Canada}
\centerline{K7L 3N6}
\vskip 2 true cm
\noindent Abstract:It is shown that, the only static asymptotically flat non
extrema black hole solution of the Einstein-conformally invariant scalar
field equations having the scalar field bounded
on the horizon, is the
  Schwarzschild one.Thus black holes cannot be endowed with conformal
scalar hair of finite length.
\vfill
\eject
\centerline{{\bf INTRODUCTION }}
To a physicist, knowing the spectrum of asymptotically flat black hole
solutions of the Einstein field equations it has been  always of a vast
 interest.
 The classical black hole uniqueness theorems assert
 that as far as the Einstein-Maxwells
equations are concerned, the only
asymptotically flat, black hole equilibrium states are
those and only those
included
 in the Kerr-Newman family of metrics.
  It is not clear however, what the state
of affairs is whenever other field configurations are
considered as sources of Einstein equations.
Further,whereas some time ago relativists
had faith in the so called "no hair conjecture", ie in
the statement that black holes
always have their exterior empty, apart from an
 electromagnetic field,
 the conjecture appears now days to faint away.
The black hole
solution discovered in [1] implies that a dilaton coupled to a
Maxwell field can withstand
the inward pull of gravity and thus peacefully
 coexists with the black hole.On the other hand the recently analyzed [2]
Einstein
non Abelian Yang-Mills equations shows that a non trivial
Yang-Mills field is a genuine hair and thus the
"colored" class of black holes
provides another counterexample against the "no hair conjecture".For the
moment however
there is no theory telling us which sources admit hairy black
holes
and which do not.In the absence of such theory
 one has to consider explicitly various energy-
momentum tensors as source of gravity and try to detect the nature of the
admitted black
holes.

The purpose of the present paper is to look for black holes solutions
admitted by the
coupled Einstein-conformally invariant scalar field equations.
Although it is
well
known [3] that a minimally coupled to
gravity massless scalar field taken as a source of Einsteins gravity
 admits only
the Schwarzschild black hole
as a solution,it is not yet known whether the same holds true
for the conformally invariant system.
On the contrary the conformal
system does admit a black hole solution.It discovered by Bekenstein [4]
and it shares the following properties:

1a) It is spherically symmetric and extrema

1b) The scalar field is unbounded on the horizon.

Despite its shortcomings (see [5] but also [4])
the existence of the "Bekenstein-black hole"
raises a number of related questions.For instance:
Is the Bekenstein black hole unique?
Do there exist other black holes solutions of the system where the field
obeys different boundary conditions on the horizon?
If they do exist,are all of them spherical,or (and)
extrema?
It is the  purpose of the present paper to investigate
some of the above questions.More precisely
 we shall search for black holes solutions which are:

2a) non extrema

2b) the conformal scalar field is bounded on the horizon.

 It will be shown in the following sections that there is only one
class of black holes states satisfying (2a,2b).
More precisely it will be proven
that the only static
asymptotically flat black hole solution obeying
 properties (2a,2b)
  is the
Schwarzschild one having vanishing exterior conformally invariant scalar
field.The result
will be shown in two different but complementary to each other
methods.
In the first approach we look for black hole states with the property
  that the non
 trivial scalar field $\Phi$
is bounded on the horizon ie
$\Phi^{2}<{\alpha}^{-1}$,
with ${\alpha}$ a constant to be determined later on.In
this case
we shall exploit the local
equivalence of the conformal system to the minimal one [6] and
this fact will allows us to prove
 a stronger statement.
Namely
 the absence of multipole static black hole endowed with
 conformal
hair.
The second method removes the restriction upon the
scalar field.In that sense it is more general than the first
one but it restricts the topology of the event horizon
(and in that sense the results are not as general as the first).
Specificaly it will be assumed that
the event horizon consists only of a single connected component.
Under this hypothesis we show that any static black hole
solution  must have vanishing scalar
hair and thus be Schwarzschild.
 Although the connectedness assumption
of the horizon is likely to be irrelevant,
 our method of
establishing the proof needs it.

 We shall begin the paper
by first giving a new prof of the black hole uniqueness theorem for the
minimally coupled system.The proof
utilizes
 a theorem proven by Shoen and Yau [9]
   and follow arguments similar
to those employed by
 Bunting and ul-Alam [7] and others [8] in re-deriving
known black hole uniqueness theorems.As a byproduct
 we also show that the the Einstein-minimal coupled massless scalar
field equations
do not admit  multipole
black hole in static equilibrium.
The main results for the conformal case are
derived in section two and three and we finish the paper with a
discussion related to the
uniqueness of the Bekenstein black hole.
\vskip 0.5in
\centerline{{\bf 1 Uniqueness Theorem for the minimal system.}}

Some time ago, Chase and Bekenstein [3] established the following
  theorem:The only static, asymptotically
flat, black hole solution of the Einstein-minimally coupled scalar
field equations is the Schwarzschild black hole possessing a trivial
(ie vanishing) exterior
   field.
 We shall present a new prof of the above theorem
 the results of which,
will serve as an
intermideate step in reaching our goal for the conformal
coupling.

We begin by introducing a set of coordinates suitable for descriding
 static black hole exterio.
 The presence of the hypersurface
orthogonal Killing field allows us to
 introduce coordinates
so that the line element in that part of the spacetime reads:
$$ds^2=-V^2dt^2 + g$$
 $V^2$ stands for the square of the timelike Killing field and $g$
is a Riemmanian metric on the $t=Const$ spaces.
In the above set of coordinates, the coupled Einstein-minimal
scalar field
equations
 are equivalent to:
$$R_{\alpha\beta}=kD_{\alpha} \Phi D_{\beta}\Phi
 +V^{-1}D_{\alpha}D_{\beta}V\eqno(1a)$$
$$D^{\alpha}D_{\alpha}V=0\eqno(1b)$$
$$D^{\alpha}D_{\alpha}\Phi=-V^{-1}D^{\alpha}VD_{\alpha}\Phi\eqno(1c)$$
where the Ricci, all covariant derivatives,raising and lowering indices
are in terms of the positive definite metric of $t=const$ slice.Any
 $t=const$ slice shall be refered here after by $\Sigma$,while the
gravitational coupling constant is
denoted by k.
The three metric $g$ and the field $\Phi$
are assumed to exhibit the familiar fall of rate
 at the asymptotic region of
$\Sigma$ ie:
$$g_{\alpha\beta}=(1+2mr^{-1}){\delta}_{\alpha\beta}+
h_{\alpha\beta}\eqno(2a)$$
$$V=1-2mr^{-1}+u\eqno(2b)$$
$$\Phi=er^{-1}+f\eqno(2c)$$
where the functions h,u,and f are assumed to be $O(r^{-2})$ possessing
 $O(r^{-3})$
derivatives.
On the other hand the internal boundary $bd\Sigma$ of $\Sigma$ defined
by $V=0$ will de assumed to be union of regular closed two spaces
on which
the four dimensional scalar $K=R^{MNKA}R_{MNKA}$ should be bounded.
It is appropriate at this point to recall the Shoen-Yau theorem [9].It
states that an asympotically flat,complete, orientable Riemannian manifold
$(M,{\gamma})$ possesing zero ADM mass,and non negative scalar curvature,must
be flat.This conclusion constitutes the backbone for new elegant proofs
of various black hole uniqueness theorems [7,8].The idea behind the new
proofs is to
 consider first two copies of $\Sigma$ say
$\Sigma_{-}$,$\Sigma_{+}$ and introduce Riemannian metrics $g_{-}$,
$g_{+}$ on $\Sigma_{-}$ and $\Sigma_{+}$ respectively.
The first one posseses
zero ADM mass, as measured from the asymptotic region of $\Sigma_{+}$,
while $g_{-}$ exhibits asymptotic properties
guaranteeing that if one add
 a point $P$, standing for the infinity of $\Sigma_{-}$
 then,
 $M =\Sigma_{-}U(bd\Sigma_{\pm})U\Sigma_{+}$ is
a complete,non singular without boundary Riemannian manifold.
 The authors of [7] construct the metrics $g_{\pm}$ by
conformally transforming
 the physical metric $g$ of $\Sigma$.
By an appropriate choice of the conformal factor $\Omega$
one may be able to
meet all the conditions required by the Shoen-Yau theorem.
 For our problem we choose $\Omega_{\pm}={1\over4}(1 \pm V)^{2}$
 and consider $g_{\pm}=\Omega_{\pm}^{2}g$,
 as the two Riemannian metrics on
$\Sigma_{\pm}$ respectively.By construction the ADM mass of
$g_{+}$ is zero,and the asymptotic
behaviour of $V,g$, implies that
by additing a point $P$ ,$g_{-}$ compactifies the infinity of
$\Sigma_{-}$.(For more details about the construction the reader may consult
ref[7].)
Furter,one may easily prove that $g_{\pm}$
join smoothly along the two "spheres" defined by $V=0$.
 and the scalar
curvature of $g_{\pm} $ reads:
$$R[g_{\pm}]=k\Omega^{-2}D^{\alpha} \Phi D_{\alpha}\Phi\eqno(3)$$
which manifestly is not negative.
Thus according to the Shoen-Yau
theorem $g_{\pm}$  must be flat which in turn
   implies (according to the above formula) that the
 gradient of $\Phi$ must be zero.Therefore
the scalar field is constant over $\Sigma_{+}$ and
because it vanishes at infinity it must vanish everywhere.
To complete the proof and
thus show that the physical metric $g$
 is spherical and more precisely
 Schwarzschild one uses the conformally flat nature of $g$
 and appeals to the familiar arguments [10].
(Although
we have never
mentioned it explicitly,
it might be worth keeping in mind that the proof requires the
gradient of $\Phi$ to be bounded on the horizon.Otherwise the curvature
of $g_{\pm}$ will be singular and thus one cannot appeal to Shoen-Yau theorem.)
\vskip 0.5in
\centerline{\bf 2 The conformal system with $\Phi^{2}<{\alpha}^{-1}$}

Let us now move to our main problem ie the examination of the coupled
Einstein-
Conformally invariant scalar field equations.The theory is described by the
following set of equations:
$$R_{MN}=({{\alpha} \over 1-\alpha\Phi^{2}})[4\nabla_{M} \Phi
\nabla_{N}\Phi
-2 \Phi\nabla_{M}\nabla_{N}\Phi-
g_{MN}\nabla^{\Gamma} \Phi\nabla_{\Gamma}\Phi]\eqno(4a)$$
$$\nabla^{A}\nabla_{A}\Phi=0\eqno(4b)$$
where all indices are four dimensional and $\alpha=k6^{-1}$.
 The system
is much more complicated than say the minimally coupled one.
The "conformal" stress tensor contains
second derivatives of the scalar field
and this
fact introduces additional complications demanding special care.
 Even looking for exact solutions is a
rather difficult and painful job [11].
Our task is to investigate whether the coupled system admits
asymptotically flat black hole
states obeying conditions (2a,2d).Let us assume that
it does so.At first note that
the non extrema property of the horizon implies
 non vanishing surface
gravity and the $t=cons$ slices smoothly intersect
the
bifurcation two spheres [12].
Projecting the above equations along the $t=cost$ spaces one get
an
equivalent set:
 $$R_{\alpha\beta}={\alpha\Lambda}[4D_{\alpha} \Phi D_{\beta}
 \Phi-2\Phi D_{\alpha}D_{\beta}\Phi
-g_{\alpha\beta}D^{\gamma}\Phi D_{\gamma}\Phi]
+V^{-1}D_{\alpha}D_{\beta}V\eqno(5a)$$
$$V^{-1}D^{\alpha}D_{\alpha}V={\alpha\Lambda}[{D^{\alpha} \Phi D_{\alpha}\Phi
+2\Phi V^{-1}D^{\alpha} \Phi D_{\alpha}V}] \eqno(5b)$$
$$D^{\alpha}D_{\alpha}\Phi=-V^{-1}D^{\alpha} \Phi D_{\alpha}V \eqno(5c)$$
where we denote here after:
$$\Lambda={1\over{1-{\alpha}\Phi^{2}}}\eqno(5d)$$

 The coupled equations can be transformed to a
 set
 describing a self
gravitating
  minimally coupled scalar field.Such a reduction
 has been known for
some time  [6,13] and it has been used as a
 mechanism to generate solutions of the conformal system from that of the
minimal one.The reduction can be accomplished via a
 conformal transformation of the metric and
simultaneous field redefinition.To benefit from this property of the conformal
system one must search for a smooth
 conformal factor which:

3a) should not vanish neither within $\Sigma$ nor
on $bd\Sigma$.

3b) should transform the conformal black hole solution
to an asumptotically flat solution of the minimal sustem.

 For at least one case such a conformal factor (actually an equivalence
class)
can be constructed
explicitly.Before we do so,
let us introduce a new metric via:
$$\overline{g}=(1-{\alpha}\Phi^{2})g\eqno(7)$$
and simultaneously define a new red shift factor $\overline{V}$ and field
$\overline{\Phi}$ by:
$$\overline {V}=(1-{\alpha} \Phi^{2})^{1\over2}V\eqno(8)$$
$$\Phi={\alpha}^{-{1\over 2}}
tanh({\alpha}^{1\over 2} \overline{\Phi}\eqno(8a)$$

It takes
 some calculations to show that because of (5a-c)
$\overline {g}$, $\overline{V}$ and $\overline{\Phi}$
obey:
$$R[\overline{g}]=k\overline{D}^{\alpha}\overline{\Phi} \overline{D}_{\alpha}
\overline{\Phi}\eqno(9a)$$
$$\overline{D}^{\alpha}\overline{D}_{\alpha}\overline{V}=0\eqno(9b)$$
$$\overline{D}^{\alpha}\overline{D}_{\alpha}\overline{\Phi}=-\overline{V}
\overline{D}^{\alpha} \overline {\Phi}
\overline{D}_{\alpha} \overline{\Phi}\eqno(9c)$$
 A comparison between the above set of equations and (1a-c)
 shows that the barred quantities indeed
satisfy field equations
 identical to those obeyed
by the minimally coupled system.Therefore
 any
 conformally transformed black hole solution is mapped locally
to solution of the minimal system.However that by itself allows us to
conclude almost
nothing.We must demonstrate
 that the conformal factor $\Omega^{2}=
1-{\alpha}\Phi^{2}$ is a "good" one, ie
 obeys
properties (3a,3b).A-priori however there is no fundamental
reason that the above defined
 $\Omega$ should  satisfy them.
It is at this point where restrictions upon $\Phi$ will be imposed.
Notice that by applying maximum principle to the
equation (5c) one concludes that $\Phi$ cannot have neither maximum nor
minimum
in the interior of
$\Sigma$.
Thus $\Phi$ is monotonic over $\Sigma$
and consequently if the conformal factor is positive on the horizon it
will remain so within the interior of $\Sigma$.This will be
the case provided
on the horizon [13] the  field satisfies $\Phi^{2}<{\alpha}^{-1}$.
In such
case it is clear that our choice of the
conformal factor globally maps the conformal black hole solution to a
coressponding black hole of the minimal system.
However because the latter system admits only one class of black holes
namely the Schwarzschild class,it is clear from (8a)
 that our initial conformal black hole solution
 must have vanishing
scalar
field and further be spherical
ie be the Schwarzschild black hole.
Notice again, that the proof requires
 the gradient of $\Phi$ to be finite on the horizon,a point that will be
discussed at length latter on.
\vskip 0.5in
\centerline{\bf 3 The case with $\Phi^{2}\geq {{\alpha}^{-{1}}}$}

Although it was relatively easy to rule out the existence of
hairy black holes with the property $\Phi^{2}<{\alpha}^{-{1}}$
ruling out the existence of
 black holes obeying $\Phi^{2}\geq {{\alpha}^{-{1}}}$
over the horizon
 it is not straightforward.
 This is due to the fact that the obvious cice
$\Omega^{2}={\alpha}\Phi^{2}-1$ would not any longer do
the job.One may
prove that the conformal solution
is mapped onto a non asymptotically flat solution of the minimal system
[6b] a conclusion that leads us to nowhere.We shall circumvented the difficulty
 by
 employing an alternative method.Unfortunately however, at some cost.Whereas so
far
we let the horizon to be arbitrary union of two "spheres" corresponding to the
 notion of multiple black holes, from now on some restriction will be imposed.
In particularly  we shall
assume that the horizon consists of a single
connected component.In turn, this will allows us
to use some of the machinery developed by Israel [14] in establishing the
uniqueness of the Schwarzshild black hole.
Recall in Israels approach one uses $V$ as one of the coordinates
and projects the field equation on each $V=cons$ two space.
They naturally split into
a set of dynamical equations propagating the intrinsic metric $g_{ij}$
extrinsic curvature $K_{ij}$, and fields from one two space into next one
and a set of constraints equations relating the dynamical variables on some
initial surface.For our purpose we shall explicitly use
 only the constraint system.
An insightful tool for our subsequent analysis is the
 four dimensional scalar $R^{ABKM}R_{ABKM}$. In Israels approach takes
the form:
$${1\over4}R^{ABKM}R_{ABKM}=G_{\alpha\beta}G^{\alpha\beta}+{K^{ij}K_{ij}
\over{\rho}^{2}V^{2}}+{2D^{i}{\rho}D_{i}{\rho}\over{\rho}^{4}V^{2}}
+{1\over{\rho}^{2}V^{2}}
 [{{k{\rho} V}\over2}{(\mu +^{3}T)+K)}]^{2} \eqno(6a)$$
where $G_{\alpha\beta}$ is the three dimensional Einstein tensor,
 $K$ is the trace of the extrinsic
curvature of the $V=const$ equipotential
two-spaces and ${\rho}^{-2}=D{\alpha} VD_{\alpha} V$.
The scalars ${\mu},^{(3)}T$ are contractions of the conformal stress
tensor and are giving by:
$${\mu}=T_{MN}n^{M}n^{N}={1\over(1-{\alpha}\Phi^{2})}
[{2\Phi\over V} D^{\alpha} V D_{\alpha} \Phi+D^{\alpha} \Phi D_{\alpha}
\Phi]\eqno(6b)$$
$$^{3}T=T_{\alpha\beta}g^{\alpha\beta}=
T_{MN}{h_{\alpha}}^{M}{h_{\beta}}^{N}
g^{\alpha\beta}={1\over 1-\alpha\Phi^{2}}[D_{\alpha} \Phi
D^{\alpha} \Phi+{{2\Phi}\over V} D^{\alpha}\Phi
D_{\alpha} V]\eqno(6c)$$
where $n^{A},{h_{\alpha}}^{M}$ are the unit normal of the $t=const$ slice
and the projection tensor on the slice respectively.There
are  two particular places where $R^{ABMN}R_{ABMN}$ pics up peculiar
contributions due to
the conformally invariant nature of $\Phi$.
First in the combination ${\mu}+^{3}T$ and
secondly in the three dimensional Einstein tensor where second derivatives of
$\Phi$ are appearing.Both of them have to be regular on the horizon.
Let $g,V,\Phi$ describes a black hole solution subject
$\Phi\geq{{\alpha}^{-{1\over 2}}}$.(Although this is the case we
like to investigate,it ought to be stressed that the following proof is
 independent upon the particular value of $\Phi$ on the horizon).
We shall first show that if $\Phi$ is bounded on
 the horizon, then there it must be
 constant with vanishing three
gradient.The claim is a consequence of the regularity of
$R^{ABMN}R_{ABMN}$ on the horizon.
Regularity of
the right hand side of (6a) on $V=0$,
implies that $V=0$ is a
totally geodesic two-space on which $D_{i}\rho=O({\rho}^{2}V)$.
Further taking into account (6a,c) one concludes that
 the combination ${\mu}+^{3}T$ is finite on the horizon,
provided the expression :
$${\alpha\over1-{\alpha}\Phi^{2}}{{2\Phi}\over V}D^{\alpha}V
D_{\alpha}\Phi$$
is bounded there.Since extremal
 horizons have been excluded,one
therefore one concludes that the above term makes a regular contribution
provided
one of the following alternatives holds:

4a) The field $\Phi$ is unbounded on the horizon.For instance
if  $\Phi=O(V^{-1})$
and  $D^{\alpha}\Phi D_{\alpha}V$ is bounded then everything is fine on
the horizon [15].

4b) The field $\Phi$ is bounded but $D_{\alpha}\Phi=O(V)$

As we have already discussed in the introduction alternative 4a) has been
excluded from further considerations so attention will be will be
restricted to the second case.
But condition (4b) implies that $\Phi$ is constant
over the horizon say $\Phi=\Phi_{o}$.
Finally to conclude the regularity of the entire right handside of (9a) over
the horizon we outght to examine the second derivatives
of $\Phi$ hiden in the three dimensional Einstein tensor (see eqs (5ab))
For points interior to $\Sigma$, $\Phi$ obeys
an elliptic equation and as long as the geometry is not singular
one expects $\Phi$ to be smooth and in fact
analytic.For horizon points however this argument  is not sufficient.
 We shall show that if (4b) holds, then the second derivatives of $\Phi$ on
the horizon are bounded.
To prove it, we shall use the constraint system of the Israel
formalism.
Adopted for our case it has the form:
$${1\over 2}(^{2}R+K^{ij}K_{ij}-K^{2})=-{\alpha}T_{\alpha\beta}
n^{\alpha}n^{\beta}+{K\over V{\rho}}\eqno(9a)$$
$$\overline {D_{c}}({K^{c}_{i}}-{{\delta}^{c}}_{i}
 K)=
{\alpha}T_{ai}n^{\alpha}
-{1\over(V\rho^{2})}\overline {D_{i}}{\rho}\eqno(9b)$$
with the components of $T_{\alpha\beta}$ giving by:
$$T_{\alpha\beta}={\Lambda}(4D_{\alpha}\Phi D_{\beta}\Phi-
2\Phi D_{\alpha} D_{\beta}\Phi-
g_{\alpha\beta}D^{\gamma} \Phi D_{\gamma} \Phi)\eqno(9c)$$
The overbarred covariant derivative differentiates tangentially and is
formed out of the induced two metric,while the reader is reminded that latin
indices are two dimensional.
Specializing the above equations to the $V=0$ space and taking into account
 (4b) one finds that the second constraint is
automatically satisfied while the first reduces to:
$${1\over 2}(1-{\alpha}\Phi^{2})
(^{2}R-{{2K}\over{V{\rho}}})
=2{\alpha}\Phi n^{\alpha}n^{\beta}D_{\alpha}D_{\beta}\Phi\eqno(10)$$
The above constraint relates $\Phi$
and its second radial derivative on the
horizon to the Gaussian curvature $^{2}R$ and extrinsic curvature.
Because the intrinsic
geometry is regular one concludes that the second radial
derivative of $\Phi$ is bounded on the horizon.The tangential
and other mixed derivatives are vanishing as a consequence of the constancy of
$\Phi$ over the horizon.
Notice that as a consequence of (10) one gets  even
stronger conclusion.The ratio
$${D_{\alpha} D_{\beta} \Phi}\over{1-{\alpha} \Phi^{2}}$$ is always bounded
on the horizon which also implies regularity even in the case where
on the horizon $\Phi^{2}={\alpha}^{-1}$.

Having in mind the above detailed boundary behaviour of
$\Phi$ on the horizon, let us multiply both sides of (5c) by $V$.It leads to
$$D^{\alpha} (V D_{\alpha} \Phi)=0$$
Multiplying further by $\Phi$,
integrating
over $\Sigma$ and applying Stokes theorem one
arrives at:
$$\oint{V\Phi D_{\alpha}\Phi d{\Sigma^{\alpha}}}
=-\int{VD^{\alpha}\Phi D_{\alpha}\Phi} d^{3}V$$
where the surface integral is taken over the $V=0$.The contribution from
the surface integral at infinity has been neglected since the asymptotic
behaviour of the geometry and $\Phi$ implies that the integrand is
$ O(r^{-3})$ at infinity.Taking into account the behaviour of
$\Phi,D_{\alpha}\Phi$ on the inner boundary we see that there is not any
contribution either.Therefore the left side of above equation is vanishing
implying  $D_{\alpha}\Phi$ is zero on ${\Sigma}$ ie $\Phi$
is identically zero in the black hole exterior.Proving further that the
system (5a-c) with $\Phi=0$
admits only the Schawrtzschild black hole as a solution is
by now well known.
\vskip 0.5in
\centerline{{\bf Discussion}}
Although we have proven the absence of non trivial static black holes
endowed with conformal hair of finite
length and possesing non degenerate horizons, the case of black holes admitting
infinite
conformal hair is still open.
For the later case,
 it is of great interest to know whether there exist non
extrema solutions.
Surely if they
do exist they cannot be spherical.This is so because of a
limited uniqueness theorem obeyed by the Bekenstein solution.
It was shown elsewhere [16]
  that amongst
all static, asymptotically flat,
spherical solutions of the system, the only black hole solution is the
(extrema) Bekenstein black hole.This result
although very useful it is limited, because it only
  examines the spherically symmetric sector of the system and does not  rule
out  static axisymmetric black hole solutions.This
seem to be unlike and all existing
evidences suggest that the only non trivial static,asymptotically
flat
 black hole solution of the Einstein-
conformally invariant scalar field eqs.is the Bekenstein black hole.At the
moment there is no proof of this statement.
It may be worth to reminder the reader that nothing is known about
existence of stationary
black hole states admitted by the conformal system.

This paper is the outcome of an active and enjoyable collaboration of the
author with the late B.C.Xanthopoulos and it constitutes a natural extension of
the results presented in ref [16].I take the oppurtunity to dedicate it to his
memory.

Part of this research was done while the author was a
 a  CITA National Fellow and was partially supported by a grant
from the Principals Development Fund of Queens University.
\vfill
\eject
\noindent

\vfill
\eject
\parindent 0pt
\centerline{{ bf\ References}}

[1] G.W.Gibbons Nucl.Phys. B207,337,(1982)\\
G.W.Gibbons and K.Maeda Nucl.Phys. B298,741,(1988)\\
G.Horowitz,A.Strominger,S.Giddings Phys.Rew.D43,3140 (1991)\\
(Err.D45,3888,1992)\\
[2] P.Bizon,Phys.Rev.Lett. 64,2844 (1990)\\
M.S.Volkov and D.V.Galtsov,Sov.J.Nucl.Phys. 51,747 (1990)\\
H.P.Kunzle and A.K.M.Masood-ul-Alam,J.Math.Phys. 31,928(1990)\\
See also: D.Sudarsky and R.M.Wald,Phys.Rev.D46,1453 (1992)\\
[3] J.E.Chase,Commun.Math.Phys. 19,276 (1970)\\
J.D.Bekenstein,Phys.Rev.D5,2941 (1972)\\
[4] J.D.Bekenstein Ann.Phys.(N.Y) 82, 535 (1974)\\
Various properties of the solution are discussed in:\\
J.D.Bekenstein Ann.Phys.(N.Y.) 91,72. (1975)\\
[5]The solution however appears to be unstable see for instance:\\
N.Bocharova,K.Bronnikov and V.Melnikov Vestn.Mosk.Univ.Fiz.Astron.\\
{}~~~~6,706, (1970)\\
[6]See J.D.Bekenstein in ref [4],also\\
B.C.Xanthopoulos and A.DialynasJ.Math.Phys.J.Math.Phys.33,1463,(1992)\\
[7]G.L.Bunting and A.K.M. Masood-ul-Alam Gen.Rel.Grav.19,174 (1987)\\
[8]  A.K.M Masood-ul-Alam Class.Quantum Grav. 9,L53 (1992)\\
{}~~~~ M.Heusler Class.Quantum Grav. 11 ,L49 (1994)\\
{}~~~~P.Ruback Class.Quantum Grav.  5 ,L155 (1988)\\
[9]  R. Schoen and S.T. Yau, Commun. Math Phys. 65,45 (1979),also\\
{}~~~~ R.Shoen and S.T.Yau,Commun. Math Phys. 79, 231, (1981),\\
[10]D.C.Robinson G.R.G 8,695,(1977) \\
see also L.Lindblom J.Math.Phys.21,1455,(1980)\\
[11] B.C. Xanthopoulos and T.Zannias J.Math. Phys. 32,1875 (1991)\\
{}~~~~B.C. Xanthopoulos and T.Zannias J.Math.Phys. 33, 1462 (1992)\\
[12]I.Racz anb R.M.Wald Class.Quantum Grav.9,2643,(1992)\\
[13]It ought to be reminded that the self gravitating coupled system
it is not even
scale invariant.Thus if $(g_{MN},\Phi)$ is a solution of the system
 one cannot alter the value of $\Phi$ on the
horizon by a rescaling of the form $c^{2}g_{MN},c^{-1}\Phi$ with $c$ an
arbitrary constant.The only case one can do that is via rescaling of the
form $(g_{MN},c\Phi)$ and simultaneously renormalizing the coupling
constant to a new value $kc^{-2}$.However in this case the numerical value
of $1-{\alpha}\Phi^{2}$ remains invariant.\\
[14]W.Israel Phys.Rev 164,1776(1967) and Comm.Math.Phys. 8,245,(1968)\\
[15]Actually this case requires a rather delicate balance
between the way $D^{a}\Phi D_{a}\Phi$ and $\Phi$ approach infinity.\\
[16]B.C.Xanthopoulos and T.Zannias J.Math.Phys.32,1875,(1991)\\

\bye